\documentclass[11pt, draftclsnofoot, onecolumn, letterpaper]{IEEEtran}

\usepackage{graphicx}

\usepackage{subfigure}
\usepackage{cite}
\usepackage{graphicx}

\usepackage{amsmath}

\usepackage{pgfplots}
\usepackage{lettrine}
\usepackage{siunitx}
\usepackage{amssymb, bbm}
\usepackage{amsthm}
\usepackage{mathrsfs}
\usepackage{mathtools}
\usepackage[algoruled,boxed,lined]{algorithm2e}
\usepackage{algpseudocode}
\usepackage{psfrag}
\usepackage[margin=1in,dvips]{geometry}
\usepackage[ulem=normalem]{changes}

\let\oldbrace\{
\def\{{\oldbrace\kern0.5pt}

\newcommand{\Cc}{\mathcal{C}}

\newcommand{\Jc}{\mathcal{J}}

\newcommand{\Nc}{\mathcal{N}}

\newcommand{\Vv}{\boldsymbol{V}}

\let\P\relax
\DeclareMathOperator\P{\sf P}

\DeclareMathOperator\C{C}

\newcommand{\U}{\mathrm{Unif}}

\newcommand{\Real}{\mathbb{R}}

\newcommand{\sfrac}[2]{\mbox{\small$\displaystyle\frac{#1}{#2}$}}

\usepackage{tikz,tkz-graph}
\usepackage[mode=buildnew]{standalone}
\usetikzlibrary{matrix,positioning,fit,calc,shapes}
\pgfplotsset{compat=newest}
\algnewcommand{\Initialize}[1]{%
  \State \textbf{Initialize}
  \Statex \hspace*{\algorithmicindent}\parbox[t]{0.8\linewidth}{\raggedright #1}
}

\newcommand{\Ndbs}{N}
\newcommand{\Ngt}{K}

\newcommand{\Wv}{\mathbf{W}}

\newtheorem{theorem}{Theorem}
\newtheorem{lemma}{Lemma}

\newtheorem{proposition}{Proposition}

\theoremstyle{definition}

\newtheorem{remark}{Remark}

\IEEEoverridecommandlockouts

\begin{document}

\title{Cooperative Strategies for {UAV}-Enabled Small Cell Networks Sharing Unlicensed Spectrum}
\author{\IEEEauthorblockN{Yujae Song, Sung Hoon Lim, Sang-Woon Jeon, and Seungjae Baek
}

\thanks{Y. Song, S. H. Lim, S. Baek are with the ICT R\&D unit, Korea Institute of Ocean Science and Technology, Busan 49111, Korea (e-mail: yjsong@kiost.ac.kr; shlim@kiost.ac.kr; baeksj@kiost.ac.kr) }
\thanks{S.-W. Jeon is with the Department of Military Information Engineering, Hanyang University, Ansan 15588, South Korea (e-mail: sangwoonjeon@hanyang.ac.kr).}%
}%

\maketitle


\maketitle
\begin{abstract}
In this paper, we study an aerial drone base station (DBS) assisted cellular network that consists of a single ground macro base station (MBS), multiple DBSs, and multiple ground terminals (GT).
We assume that the MBS transmits to the DBSs and the GTs in the licensed band while the DBSs use a separate unlicensed band (e.g. Wi-Fi) to transmit to the GTs. 
For the utilization of the DBSs, we propose a cooperative decode--forward (DF) protocol in which multiple DBSs assist the terminals simultaneously while maintaining a predetermined interference level on the coexisting unlicensed band users. For our network setup, we formulate a joint optimization problem for minimizing the aggregate gap between the target rates and the throughputs of terminals by optimizing over the 3D positions of the DBSs and the resources (power, time, bandwidth) of the network.  To solve the optimization problem, we propose an efficient nested structured algorithm based on particle swarm optimization and convex optimization methods. 
Extensive numerical evaluations of the proposed algorithm is performed considering various aspects to demonstrate the performance of our algorithm and the gain for utlizing DBSs. 

\end{abstract}

%
%

\section{Introduction} 

\lettrine{U}{tilization} of aerial drones as mobile base stations is a new paradigm that can enhance network throughput and extend coverage~\cite{Zeng:16, Irem:16,Lav:16} in cellular networks.
Unlike static ground base stations, drone base stations (DBSs) can be dynamically deployed and can adjust their positions to support various quality of service (QoS) requirements and balance load between heterogeneous cells by providing alternative communication links in an efficient manner \cite{Antonino:17}.
In some cases, for instance, natural disasters or in marine environments, deploying ground base stations (BSs) in a timely manner can be almost impossible, and for such cases a DBS aided communication network can be extremely helpful~\cite{Moren:17}. As in these use cases, the main advantage for using drones as aerial BSs comes from their mobility and promptness, 
which provide an efficient way of establishing alternative wireless links. In order to fully utilize such wireless links, however, the positions of the DBSs have to be carefully optimized while considering the data traffic demands, or more generally, the overall QoS requirements of the heterogeneous users and services.    

To address the DBS placement problem, there has been some series of works that optimize the placement of the DBSs or drone relays when the altitude of the drones are fixed \cite{Rohde:13,Merwaday:15,Li:15,Mozaffari:15,Lyu:17}.  

Another important fundamental issue for drone aided communications is to establish an air-to-ground channel model which is critical for theoretical system engineers to develop an analytical and tractable framework for designing DBS aided networks.
For such purpose, air-to-ground path loss has been measured and modeled for both urban and suburban environments \cite{AlHourani:14,Feng:16,Wang:17,Al-Hourani:17}. It has been reported in \cite{AlHourani:14,Feng:16} that the probability that an air-to-ground channel becoming a line of sight (LoS) channel increases as the altitude of DBSs increases, which in general has better path loss compared to non-LoS channels. However, as in all wireless channels, the path loss also increases due to the increased distance between the DBSs and the ground users.
The two opposing aspects on the DBS altitude inherent a fundamental tradeoff. In \cite{Zheng:13,Al-Hourani2:14}, the optimal altitude of a DBS that maximizes its cell coverage have been established. 

While taking into account the air-to-ground path loss models, several 3D placement optimizations including the altitude of a DBS has been studied in~\cite{Yaliniz:16, Alzenad:17, Shakhatreh:17,Shakhatreh2:17,  Alzenad:18, Rozhina:18}.    
The 3D placement problem of a (single) DBS has been considered under various criterions, e.g., for maximizing the number of serviced users that guarantee a certain path loss threshold~\cite{Yaliniz:16}, for minimizing the transmit power of the DBS~\cite{Alzenad:17}, and for supporting different QoS requirements~\cite{Alzenad:18}.   
In \cite{Shakhatreh:17,Shakhatreh2:17}, an efficient 3D placement algorithm for a single DBS has been proposed for indoor users inside a high-rise building based on a realistic outdoor--indoor path loss model. Reinforcement learning has been applied in~\cite{Rozhina:18} to guarantee a QoS by dynamically adjusting the 3D location of a DBS based on the users mobility.

Expanding the scope to multiple DBSs~\cite{Shah:17, Elham:16, Azade:17}, distributed algorithms that associate ground small BSs along with DBSs have been proposed in~\cite{Shah:17} to maximize the overall sum rate. A 3D placement optimization of multiple DBSs has been studied in \cite{Elham:16}, in which its goal was to find the minimum number of DBSs and their 3D placement such that all users are served. In \cite{Azade:17}, dynamic DBS repositioning has been studied to improve the overall spectral efficiency by adjusting the positions of the DBSs in response to the location change of users.

An important distinction between ground BSs and DBSs is the fact that ground BSs are typically connected with the core network via wired links that have large bandwidth.
On the other hand, DBSs are connected with the macro BSs or the core network via finite-capacity wireless backhaul links, which in some cases will act as a bottleneck in the overall data transmission~\cite{Kalantari2:17}. More importantly, since the link between the marcro BS and the DBSs are wireless, the backhaul link depends on the locations of the DBSs.
Despite these clear distinctions between wired BSs and DBSs, the aforementioned literature do not take into account the location-dependent finite-capacity wireless backhaul links in the optimization of the DBSs locations.

Addressing the entirely wireless nature of the DBSs, the authors in~\cite{Kalantari:17} modeled the backhaul link as a wireless channel in which the capacity is determined by the signal-to-noise ratio (SNR) from the ground BS to each DBS, i.e., the Shannon capacity of the wireless channel. Under the general wireless model, an efficient 3D placement strategy and user-to-DBS association policy has been proposed. 

\subsection{Contributions}
In this paper, we study a DBS-assisted cellular network that consists of a single ground macro base station (MBS), multiple aerial DBSs, and multiple ground terminals.
We establish a joint optimization problem to minimize the aggregate gap between the target rates and the served rates of each terminal by optimizing over the 3D positions of the DBSs, bandwidth  and power allocations of the MBS, transmitting time fraction of the DBSs, and the terminal-to-DBS associations. 

While we ourselves refer to the drones as ``base stations'' on account of the term being the dominant terminology in the literature, strictly speaking, we propose to utilize the drones as {\em relays}. Indeed, utilizing the drones in the form of a baseline small cell base station can be considered as a primitive utilization of the decode--forward (DF) relaying strategy~\cite{Cover--El-Gamal1979}. In this point of view, we propose the use of a more sophisticated relaying strategy with several distinct characteristics. 

Firstly, the DBSs do not have separately dedicated ``backhaul'' links. Instead, the DBSs receives from the MBS using some portion of the total band that the MBS uses to transmit directly to the ground terminals.
Secondly, a ground terminal can be aided by multiple DBSs simultaneously, and lastly, the DBSs are assumed to transmit to the ground terminals using a separate unlicensed band. 
The full description of the DF strategy is given in Section~\ref{sec:cooperative_DF}.

In the following, we highlight some of our contributions:
\begin{itemize}
\item Unlike the previous work that assume a simple two-hop decode--forward (DF) protocol, we adopt a cooperative DF protocol utilizing multiple DBSs. Each terminal recovers its desired information by jointly decoding the channel observations from the MBS transmission (the direct link) and the channel observations from the DBSs.  
To the best of our knowledge, our work is the first result that studies a 3D drone placement optimization problem using a cooperative multi-relay DF protocol.
\item Based on the cooperative DF protocol, we establish a general framework for optimizing the 3D drone locations while incorporating backhaul constraints of the DBSs and cooperative relaying with multiple DBSs. Thus, our approach provides a comprehensive understanding on DBS location optimization, multi-relay cooperative DF strategies, and resource optimizations. 
\item To enhance bandwidth utilization, we assume that the cellular network bandwidth is exclusively reserved for the transmissions sent from the MBS, i.e., the MBS-to-DBS links and the MBS-to-terminal links. For the DBS-to-terminal links, we assume that the DBSs opportunistically transmit to its associated terminals using an unlicensed band (e.g. Wi-Fi band) while maintaining a permitted level of interference to the existing Wi-Fi participants.
We reflect such spectrum sharing constraints for the DBSs in our optimization framework.
\item We formulate an optimization problem for minimizing the aggregate gap between the target rate and throughput with respect to the DBS locations and resource allocations. To solve the optimization problem, we propose a nested structured algorithm. The outer loop follows a generic particle swarm optimization (PSO) procedure which efficiently finds a local optimum over the optimization space in a parallel manner. The inner loop, which computes the cost function of the PSO algorithm, is formulated as an optimal resource allocation problem given the three dimensional DBS locations. For this step, we rigorously prove that the problem is {\em convex} for which the optimal solution can be found using standard convex optimization methods.
\end{itemize}

\subsection{Paper Organization and Notation}
Starting with the next section, we formally present our network model, formulate the problem statement, and present the cooperative multi-DBS DF strategy. 
In Section~\ref{sec: algorithm} we present our main algorithm for optimizing the DBS locations and the resource allocations. In Section~\ref{sec:numerical_evaluation}, we present various numerical evaluations that show the performance improvements of our strategy, and finally we give some concluding remarks in Section~\ref{sec: concluding_remark}.

We denote $[a:b]$ as a sequence of integers $\{a,\ldots, b\}$, $[a,b]$ as $\{x\in\Real: a\le x\le b\}$, $\|\cdot\|$ as the Euclidean norm, and $\C(x)=\log(1+x)$.

%
%

\section{Problem Setup} \label{sec: problem_statement}

\begin{figure}[h!]
\begin{center}
\footnotesize
\psfrag{n1}[c]{Aerial DBS}
\psfrag{n2}[c]{Ground terminals}
\psfrag{n3}[c]{Ground MBS}
\includegraphics[width=0.6\textwidth]{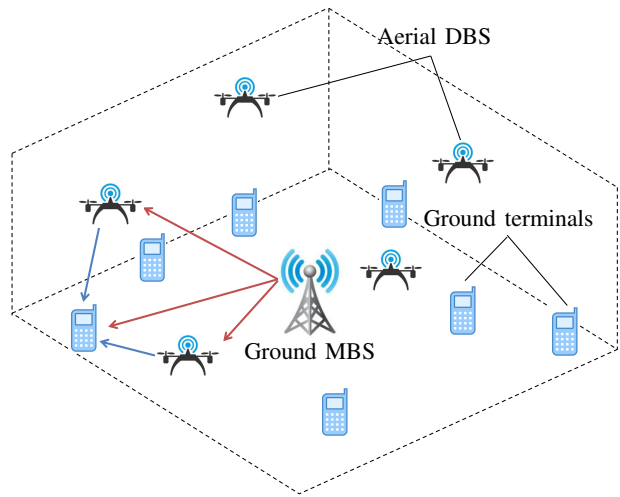}%
\caption{An Aerial DBS assisted cellular network. The network consists of a single macro base station located at the origin, multiple aerial DBSs, and multiple ground terminals. The MBS is assumed to communicate with the DBSs and the terminals in the licensed band while the DBSs transmit to the terminals using an unlicensed band. }\label{fig:network_model}
\end{center}
\end{figure}

We consider an aerial DBS-assisted small-cell cellular network which consists of a single ground MBS, $\Ndbs$ aerial DBSs, and $\Ngt$ ground terminals as shown in Figure~\ref{fig:network_model}.
Let $\phi_0=(\phi_{0,x},\phi_{0,y},\phi_{0,z})$, $\phi_j=(\phi_{j,x},\phi_{j,y},\phi_{j,z})$, and $\psi_k=(\psi_{k,x},\psi_{k,y},\psi_{k,z})$ denote the 3D positions of the MBS, DBS $j$, and terminal $k$ respectively, where $j\in[1:N]$ and $k\in[1:K]$.
Let 
\begin{align}
\mathcal{B}&=[x_{B,\min},x_{B,\max}]\times [y_{B,\min},y_{B,\max}]\times \{0\},\\
\mathcal{D}&=[x_{D,\min},x_{D,\max}]\times [y_{D,\min},y_{D,\max}] \nonumber\times [z_{D,\min},z_{D,\max}],\\
\mathcal{T}&=[x_{T,\min},x_{T,\max}]\times [y_{T,\min},y_{T,\max}]\times \{0\}
\end{align}
be the available network area for the MBS, DBSs, and terminals, respectively.
In other words, $\phi_0\in\mathcal{B}$, $\phi_j\in\mathcal{D}$, and $\psi_k\in\mathcal{T}$ for $j\in[1:N]$ and $k\in[1:K]$.

The overall network utilizes two separate bands, a license band (say, LTE) which connects the MBS with both the DBSs and the terminals, and an unlicensed band (say, Wi-Fi) which supports the connection between the DBSs and the terminals. The network is operated in two possible protocols, simply by direct communication between the MBS and terminals or by a cooperative multi-relay DF protocol~\cite{Cover--El-Gamal1979} assisted by the DBSs. 
In the following we explain how the two protocols operate. 

%
%

\subsection{Direct Transmission Protocol}
In the direct transmission protocol, the MBS sends a message $m_k\in[1:2^{nr_k}]$ to terminal $k\in[1:K]$ by frequency division multiple access (FDMA) in the license band, where $r_k$ denotes the message rate for terminal $k$.
The complex baseband channel output at terminal $k\in[1:\Ngt]$ at time slot $t\in[1:n]$ is given by
\begin{align} \label{eq:gt-L-output}
y_{k,0}[t] &= a_{k,0} x_{k,0}[t] + z_{k,0}[t],
\end{align}
where $z_{k,0}\sim\mathcal{CN}(0, N_0)$ is the complex additive white Gaussian noise (AWGN) at terminal $k$ in the license band, $a_{k,0}$ is the channel coefficient from the MBS to terminal $k$, and $x_{k,0}$ is the channel input transmitted at the MBS designated to terminal $k$ with bandwidth $b_{k0}$ and average power $p_{k0}$, i.e., $\sum_{t=1}^n |x_{k,0}[t]|^2 = np_{k0}$. We assume that the MBS is subject to a total power constraint, 
\begin{align} \label{eq:power-constraint}
\sum_{k=1}^{\Ngt} p_{k0} \le P_0,
\end{align}
and a total bandwidth constraint
\begin{align} \label{eq:bandwidth-constraint}
\sum_{k=1}^{\Ngt} b_{k0} \le B_0,
\end{align}
where $P_0$ and $B_0$ are the total power and bandwidth available at the MBS, respectively. 
For the path loss of the ground connections, i.e., the path loss for the channels $a_{k,0}$, $k\in[1:K]$, we adopt the modified Hata urban model \cite{hata} given by
\begin{equation}\label{pathloss1}
\left\{
\begin{array}{ll}
122+38\log(d_{k,0}) & \mbox{if } {d_{k,0} \ge 0.05\,{\rm{km}}},\\
122+38\log(0.05) & \mbox{if } {d_{k,0} < 0.05\,{\rm{km}} }, \\
\end{array}\right.
\end{equation}
where $d_{k,0}=\|\phi_0-\psi_k\|$ denotes the distance between the MBS and terminal $k$ in kilometers.

By the channel coding theorem for band-limited discrete memoryless Gaussian channels~\cite{Wyner1966, Gallager1968}, a rate $r_k$ via direct transmission is achievable (communicated reliably) if
\begin{align} \label{eq:direct_tx_rate}
r_k &< b_{k0}\C\left(\sfrac{|a_{k,0}|^2p_{k0}}{b_{k0}N_0}\right).
\end{align}

%
%

\subsection{Cooperative Multi-DBS Decode--Forward Protocol} \label{sec:cooperative_DF}

\begin{figure}[h!]
\begin{center}
\footnotesize
\psfrag{m1}[c]{MBS}
\psfrag{t1}[l]{Terminal $k$}
\psfrag{g1}[c]{$g_{ki}$}
\psfrag{g2}[c]{$g_{kj}$}
\psfrag{g3}[c]{$g_{kl}$}
\psfrag{d1}[c]{DBS $i$}
\psfrag{d2}[c]{DBS $j$ }
\psfrag{d3}[c]{DBS $l$}
\psfrag{h1}[c]{$h_{i0}$}
\psfrag{h2}[c]{$h_{j0}$}
\psfrag{h3}[c]{$h_{l0}$}
\psfrag{a1}[c]{$a_{k0}$}
\psfrag{b1}[c]{Broadcast}
\psfrag{b2}[c]{Multi-band reception}
\includegraphics[width=0.7\textwidth]{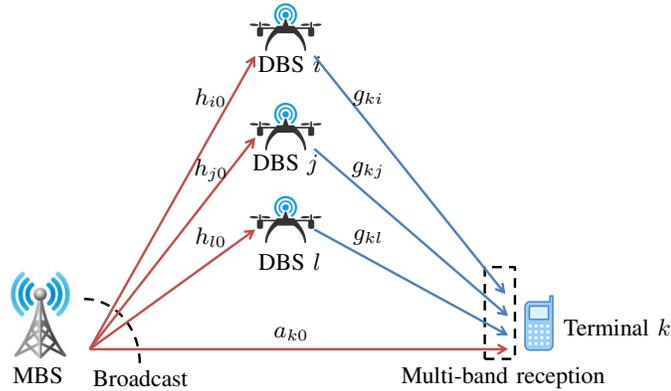}%
\caption{The cooperative DF protocol. The MBS broadcasts a codeword to the terminal and multiple DBSs (that is to assist terminal $k$'s communication) in the license band. Once the DBSs recover terminal $k$'s message, the message is re-encoded and sent to terminal $k$ using separate unlicensed bands. Note that a DBS can serve multiple terminals by further time sharing among terminal messages.}\label{fig:relay_model}
\end{center}
\end{figure}

The DF transmission strategy was first developed in~\cite{Cover--El-Gamal1979} for the three-node relay channel. Here, we propose a variant of the DF strategy which can utilize multiple DBS relays while the channel output consists of multiple orthogonal channels, i.e., we assume a multi-band receiver at each terminal to avoid interference. 
Before giving a formal description of the strategy, we briefly give an outline of the strategy. First, as in the direct transmission case, the MBS transmits the messages $m_k\in[1:2^{nr_k}]$, $k\in[1:K]$ using FDMA and a message designated for terminal $k$ is sent on the allocated licensed band. A group of DBSs recover the message for terminal $k$. Note that in the DBSs perspective, each DBS can recover multiple terminal messages. The DBSs then re-encodes each of the recovered messages and sends it to the corresponding terminal in the unlicensed band.
We note that the DBSs transmit in the unlicensed band using FDMA with equally partitioned separate bands.
If a DBS is relaying multiple messages, it additionally uses time division multiplexing between the terminal messages.
Finally, terminal $k$ recovers its message by jointly decoding over all the channel outputs sent from the DBSs and the MBS.
Since the DBSs are sending the messages via TDMA on the unlicensed band, each terminal is free from interference between the DBSs and the MBS (which transmits in the license band).
The strategy is summarized and illustrated in Fig.~\ref{fig:relay_model}.

We next give a formal presentation of the strategy. 
Let $\Jc_k\subset[1:N]$ be a set of DBS indices that assists terminal $k$.
The MBS first broadcasts a message $m_k\in[1:2^{nr_k}]$ of rate $r_k$ to DBS $j\in\Jc_k$ and terminal $k$. Then, terminal $k$ observes~\eqref{eq:gt-L-output} and DBS $j\in\Jc_k$ observes the channel outputs
\begin{align}  \label{eq:dbs-output}
y^{\operatorname{DBS}}_{j,k}[t] &= h_{j,0} x_{k,0}[t] + z^{\operatorname{DBS}}_{j,k}[t], \quad k\in[1:K]
\end{align}
where $z^{\operatorname{DBS}}_{j,k}[t]\sim\mathcal{CN}(0, N_0)$ is a complex AWGN at DBS $j$ on the band carrying $m_k$, $h_{j,0}$ is the channel coefficient from the MBS to DBS $j$. Here, $x_{k,0}$ is the channel input carrying the message $m_k$ that is broadcast to terminal $k$ and DBS $j\in\Jc_k$ with bandwidth $b_{k0}$ and power $p_{k0}$ which are subject to the constraints~\eqref{eq:power-constraint} and~\eqref{eq:bandwidth-constraint}, respectively.

To explain the DBS operation, we define time sharing parameters $\tau_{kj}$, $k\in[1:K]$, $j\in[1:N]$. The time sharing parameters $\tau_{kj}\in[0,1]$ represent the fraction of time that DBS $j$ utilizes to send terminal $k$'s designated message.
Note that this results in the constraint,
\begin{align}\label{eq:standalone-tao-constraint}
\sum_{k=1}^{K} \tau_{kj} \le 1, \quad j\in[1:N].
\end{align}
One important usage of the time sharing parameter $\tau_{kj}$ is that it also indicates which DBS serves which terminal, i.e., if $\tau_{kj}=0$, DBS $j$ does not decode nor does it relay terminal $k$'s message.
On the other hand, if $\tau_{kj}>0$, DBS $j$ recovers terminal $k$'s message, re-encodes it, and sends it to terminal $k$ using $\tau_{kj}$ fraction of its time. 

Upon receiving the channel outputs, DBS $j\in\Jc_k$ attempts to recover the message $m_k$ which can be done reliably if
\begin{align}
&r_k < b_{k0}\C\left(\sfrac{|h_{j,0}|^2p_{k0}}{b_{k0}N_0}\right), 
\end{align}
for $k\in[1:K]$ such that $\tau_{kj}\neq 0$.
After DBS $j$ recovers the allocated messages, i.e., ${m_k}$ such that $\tau_{kj}> 0$ for all $k\in[1:K]$, it re-encodes the messages and sends them to each of the corresponding terminals.
We assume that all DBSs use bandwidth $b$ and power $p$ for transmission.
At terminal $k\in[1:K]$, the observed channel outputs in the unlicensed band at time slot $t\in[1:n]$ is given by\footnote{For simplicity, we omit the time-sharing index in the output channel output model.}
\begin{align} \label{eq:gt-UL-output}
y_{k,j}[t] &= g_{k,j} x^{\operatorname{DBS}}_{k,j}[t] + z_{k,j}[t], \quad j\in[1:N]
\end{align}
where $x^{\operatorname{DBS}}_{k,j}[t]$ is the channel input transmitted from DBS $j$ to terminal $k$ with bandwidth $b$ and power $p$, $z_{k,j}[t]\sim\mathcal{CN}(0, N_0)$ is the complex AWGN at DBS $j$ on the unlicensed band, and $g_{k,j}$ is the channel coefficient from DBS $j$ to terminal $k$.

The path loss for the DBS channels $h_{kj}$ and $g_{kj}$ follow the air-to-ground model in~\cite{Air-to-Ground-channel}.
In particular, the path loss is expressed as an averaged value between line-of-sight (LoS) and non-LoS components. 
The probability of having an LoS connection between a ground entity (MBS or terminal) and DBS $j$ is given by
\begin{align*}
\P_{\operatorname{LoS}} = \frac{1}{1+\alpha\exp\left(-\beta\left(\frac{180}{\pi}\arctan\left(\frac{\phi_{j,z}}{\hat d}\right)-\alpha\right)\right)},
\end{align*}
where $\alpha$ and $\beta$ are constants that depend on the environment and are given in~\cite{Air-to-Ground-channel}, $\hat d$ is the distance between the ground entity and the projection of DBS $j$ on the ground plane, and $\phi_{j,z}$ is the altitude of DBS $j$. Overall, the path loss is given by
\begin{align}
20\log\left(\frac{4\pi f_c  d}{c}\right) + \P_{\operatorname{LoS}} L_{\operatorname{LoS}}+ (1-\P_{\operatorname{LoS}}) L_{\operatorname{NLoS}},
\end{align}
where $L_{\operatorname{LoS}}$ and $L_{\operatorname{NLoS}}$ are the path loss for LoS and NLoS also given in~\cite{Air-to-Ground-channel}, $c$ denotes the speed of light, and $f_c$ denotes the carrier frequency. 

\begin{theorem}[Cooperative DF achievable rates] \label{thm:DF}
Consider the DBS-assisted cellular network and assume that the MBS allocates power $p_{k0}$ and bandwidth  $b_{k0}$ for terminal $k\in[1:\Ngt]$. 
By the cooperative multi-DBS DF protocol, a rate $r_k$, $k\in[1:K]$ is achievable if
\begin{align} \label{eq:DF-rate}
r_k &\leq b_{k0}\C\left(\sfrac{|a_{k,0}|^2p_{k0}}{b_{k0}N_0}\right) +  \sum_{j=1}^{N}\tau_{kj}b\C\left(\sfrac{|g_{k,j}|^2p}{bN_0}\right)
\end{align}
such that 
\begin{align} \label{eq:DF-rate2}
\tau_{kj}r_k &\le \tau_{kj} b_{k0}\C\left(\sfrac{|h_{j,0}|^2p_{k0}}{b_{k0}N_0}\right), \quad j\in[1:N].
\end{align}
\end{theorem}
\begin{IEEEproof}
The proof is deferred to Appendix~\ref{app:proof-thm1}.
\end{IEEEproof}

\begin{remark}\label{rmk:universality}
We remark that the rate expressions~\eqref{eq:DF-rate} and~\eqref{eq:DF-rate2} readily includes direct transmission. This can be shown by setting $\tau_{kj}=0$ for all $j\in[1:N]$ which simplifies~\eqref{eq:DF-rate} and~\eqref{eq:DF-rate2} to~\eqref{eq:direct_tx_rate}.
\end{remark}

%
%

\subsection{Coexistence with Unlicensed Band Users}
In this paper, we include the following practical consideration when utilizing the unlicensed bands. 
We assume that each DBS shares an unlicensed band with multiple Wi-Fi access points (APs), whereas different DBSs use different unlicensed bands to avoid strong co-channel interferences.
More specifically, DBS $j$ and $M_j$ Wi-Fi APs coexist on the unlicensed band allocated to DBS $j$ (simply referred to as `unlicensed band $j$' in the sequel). We assume that the DBS adopts a listen-before-talk mechanism with random backoff using a fixed size contention window~\cite{LAA}, and that the Wi-Fi APs follow a typical CSMA/CA mechanism \cite{wifi}. 
We denote $\delta _W^{(j)}$ and $\delta _D^{(j)}$ as the channel access probabilities of the Wi-Fi APs and DBS $j$ sharing unlicensed band $j$, respectively, and $c_W^{(j)}$ and $c_D^{(j)}$ as the collision probabilities of the Wi-Fi APs and DBS $j$, respectively.
Following the steps of \cite{LAA}, the channel access probabilities of $\delta _W^{(j)}$ and $\delta _D^{(j)}$ in a randomly selected time slot are given as follows:
\begin{align}
\delta _W^{(j)} &= \frac{{2\left( {1 - 2c_W^{(j)}} \right)}}{{\left( {1 - 2c_W^{(j)}} \right)\left( {\Omega + 1} \right)c_W^{(j)}\Omega\left( {1 - {{(2c_W^{(j)})}^m}} \right)}}, \label{access_wifi} \\
\delta _D^{(j)} &= \frac{{\frac{1}{\Gamma}c_D^{(j)}\sum\limits_{i = 1}^\Gamma {{{\left( {1 - c_D^{(j)}} \right)}^{i - 1}}} }}{{1 - \frac{1}{\Gamma}\left( {1 - c_D^{(j)}} \right)\sum\limits_{i = 1}^\Gamma {{{\left( {1 - c_D^{(j)}} \right)}^{i - 1}}} }}, \label{access_DBS}
\end{align}
where $\Omega$ is the minimum contention window (CW) size of the Wi-Fi APs, $m$ is the maximum backoff stage of Wi-Fi APs, and $\Gamma$ is the backoff CW size of the DBSs. Note that the CW size of the DBSs can be adaptively adjusted to control the channel access probability of the DBSs as in~\cite{LAA}.
The collision probability of $c_W^{(j)}$ and $c_D^{(j)}$ are presented as follows:
\begin{align}
c_W^{(j)} &= 1 - {\left( {1 - \delta _W^{(j)}} \right)^{{M_j} - 1}}\left( {1 - \delta _D^{(j)}} \right), \label{collision_wifi} \\
c_D^{(j)} &= 1 - {\left( {1 - \delta _W^{(j)}} \right)^{{M_j}}}. \label{collision_DBS}
\end{align} 
Note that \eqref{access_wifi} to \eqref{collision_DBS} construct a nonlinear system of equations with four unknowns, i.e., $\delta _W^{(j)}$, $\delta _D^{(j)}$, $c_W^{(j)}$, and $c_D^{(j)}$, which can be easily solved by standard numerical methods.
Using the above results, we can determine the achievable time fraction on the unlicensed band for DBS $j$ when competing with $M_j$ Wi-Fi APs as
\begin{align}
S_D^{(j)} = \delta _D^{(j)}{\left( {1 - \delta _W^{(j)}} \right)^{{M_j}}}.\label{time_share}
\end{align}  

Each DBS timeshares in the unlicensed band with its associated users as well as the coexisting Wi-Fi APs. Note that (\ref{collision_wifi}) implies that the collision probability of Wi-Fi APs is a monotonically increasing function of the channel access probability $\delta_D^{(j)}$. In other words, higher channel access probability of the DBSs results in a higher collision probability of the Wi-Fi APs, which eventually leads to Wi-Fi performance degradation. Therefore, we can say that the collision probability $c^{(j)}_W$ measures the damage caused by the DBSs on the Wi-Fi APs. For the DBSs to coexist with the Wi-Fi APs in harmony, we restrict DBS $j$ to satisfy a collision probability constraint when transmitting on unlicensed band $j$, namely, we impose the constraint
\begin{align}
c_W^{(j)} < {C_j},
\end{align} 
where ${C_j}$ is the desired threshold for the collision probability occurred at the Wi-Fi APs on unlicensed band $j$.


\subsection{3D Placement Optimization of DBSs}
Let $r_T$ be the target communication rate  for each terminal.
The main objective of our problem formulation is to minimize the aggregate gap between the communication rate and the target rate. Formally, the primary problem of interest is
\begin{align}\label{eq:objective}
\min_\pi \sum_{k=1}^{\Ngt} (r_T - r_k)
\end{align}
subject to
\begin{align}
\sum_{k=1}^K b_{k0} &\le B_{0}, \label{eq:const1-MBSband}\\ 
\sum_{k=1}^K p_{k0} &\le P_{0}, \label{eq:const2-MBSpower}\\
\tau_{kj} b_{k0}\C\left(\sfrac{|h_{j,0}|^2p_{k0}}{b_{k0}N_0}\right) &\ge \tau_{kj}r_k, \forall k, j \label{eq:const4-backhaul}\\
{r_T} - {r_k} &\ge 0, \forall k \label{eq:const5-positive}\\
\sum\limits_{k=1}^K {{\tau _{kj}}}  &\le S_D^{(j)}, \forall j \label{eq:const6-wifi-ts}\\
c_W^{(j)} &\le {C_j}, \forall j \label{eq:const7-wifi-coexist}\\
\phi_j&\in\mathcal{D}, \forall j \label{eq:const8-location}
\end{align}
where $r_{k}$ is the achievable rate of terminal $k$ defined as  
\begin{align}  \label{eq:rate}
r_{k} =   b_{k0}\C\left(\sfrac{|a_{k,0}|^2p_{k0}}{b_{k0}N_0}\right) +  \sum_{j=1}^{N}\tau_{kj}b\C\left(\sfrac{|g_{k,j}|^2p}{bN_0}\right) ,
\end{align}
and $\pi$ is the collection of optimization parameters
\begin{align}
\pi=\{b_{k0}, p_{k0}, \tau_{kj}, \phi_j\}. 
\end{align}

For reference in the sequel, we will refer to the optimization problem in \eqref{eq:objective} to \eqref{eq:const8-location} as OP.
Note that we have two types of ``active'' rates to consider, direct transmission and cooperative DF rates. In particular, if $\tau_{kj}=0$ for all $j\in[1:N]$, then the constraint~\eqref{eq:const4-backhaul} becomes inactive and \eqref{eq:rate} is the same as the direct transmission rate in \eqref{eq:direct_tx_rate}.
The constraints \eqref{eq:const1-MBSband} and \eqref{eq:const2-MBSpower} are the total bandwidth and power constraint in~\eqref{eq:power-constraint} and~\eqref{eq:bandwidth-constraint}.
The constraint~\eqref{eq:const4-backhaul} is required to achieve $r_k$ in \eqref{eq:rate}; see Theorem \ref{thm:DF}. 
We impose the constraint~\eqref{eq:const5-positive} since we do not wish to allocate unnecessary resources to terminals that already achieve the target rate.  

\begin{remark}\label{rmk:cw}
By solving the nonlinear system of equations \eqref{access_wifi} to \eqref{collision_DBS} through adjusting the CW size of each DBS, we can obtain the maximum $S_D^{(j)}$ for DBS $j$ such that it guarantees that the collision probability $c_W^{(j)}$ satisfies~\eqref{eq:const7-wifi-coexist}. 
Thus, by fixing $S_D^{(j)}$ as the maximizing solution, we hereafter omit the constraint \eqref{eq:const7-wifi-coexist}. 
\end{remark}


Note that OP belongs to the class of multivariable non-linear programming (NLP) problems, and an optimal solution of a multivariable NLP problem is known to be NP-hard which may not be feasible with real-time algorithms. To efficiently solve OP, we instead present a heuristic suboptimal approach.

\section{Proposed Algorithm} \label{sec: algorithm}
In this section, we give a detailed description of our proposed algorithm. 
The overall framework of the proposed algorithm follows the general particle swarm optimization (PSO) algorithm~\cite{Kennedy--Eberhart1995, Shi--Eberhart1998}. 
However, we will need an additional optimization step for calculating the {\em cost} function within the PSO framework. 
Thus, we will explain the algorithm in two steps, first explaining the outerloop of the algorithm, and then give the description of the optimization steps for evaluating the cost function within the PSO algorithm.  

The concept of the PSO algorithm was first introduced in~\cite{Kennedy--Eberhart1995} which was inspired by swarm intelligence, social behavior, and food searching of bird flocks. The algorithm is widely used in the field of wireless communications due to its ease of implementation with only a few particles that are required to be tuned. 
In our framework, the algorithm starts by generating $L$ particles $\Wv_l\in \mathcal{D}^N$, $l \in[1:L]$ (in our case each particle is a set of all DBS positions) to form an initial population set. The matrix 
$\Wv_l$ contains the positions of the DBSs corresponding to the particle $l$ within an area of interest, i.e., 
\begin{equation}\label{eq:pariticle-def}
\Wv_l = \left[ {\begin{array}{*{20}{c}}
\Wv_{l,1}\\
\vdots\\
\Wv_{l,N}
\end{array}} \right] = \left[ 
\begin{array}{*{20}{ccc}}
\phi _{1,x}^{(l)}&\phi _{1,y}^{(l)}& \phi _{1,z}^{(l)}\\
\vdots & \vdots &\vdots \\
\phi _{N,x}^{(l)}&\phi _{N,y}^{(l)}& \phi _{N,z}^{(l)}
\end{array} \right].
\end{equation}

Then, for each particle $\Wv_l$, the PSO algorithm computes a cost $c(\Wv_l)$, and updates the 3D positions based on the costs of the particles.  
The outline of the algorithm is given in Algorithm~\ref{alg:PSO}. In the following we give some definitions of the parameters used in Algorithm~\ref{alg:PSO}.

\begin{algorithm}[h!]
    \SetAlgoLined
    \KwOut{$\Wv_G^\star$} 
        \textbf{Initialization:} \nonumber\\
Initialize $c_G^\star\leftarrow \infty$\;
    \For{$l\in[1:L]$}{
    Initialize $\Wv_l$ uniformly at random over $\mathcal{D}^N$\;
    $\Wv_l^\star \leftarrow\Wv_l$ and $c_l^\star \leftarrow c(\Wv_l)$\;
    	\If{$c_l^\star <c_G^\star$}{
    	$\Wv_G^\star \leftarrow \Wv_l^\star$ and $c_G^\star \leftarrow c(\Wv_G^\star)$\;
    	}
	\For{$j\in[1: N]$}{
		Initialize velocity term $\Vv_{l,j}$ uniformly at random over $\mathcal{V}$\;
		}
    }
    \textbf{Main Loop:}\\
    \While{termination criterion is not met}{
        \For{$l\in[1:L]$}{
        	\For{$j\in[1: N]$}{
        		$\Vv_{l,j} \leftarrow \eta \Vv_{l,j} + {a_1}{\mathbf{r}_1}\circ\left( \Wv^\star_{l,j} - \Wv_{l,j} \right) + {a_2}{\mathbf{r}_2}\circ\left( \Wv_{G,j}^{\star} - \Wv_{l,j} \right)$\;
		        $\Wv_{l,j}\leftarrow \Wv_{l,j} + \Vv_{l,j}$\;
		}
		
	\If{$c(\Wv_l) < c_l^\star$}{
		$\Wv^\star_{l} \leftarrow \Wv_{l}$ and $c^\star_{l} \leftarrow c(\Wv_{l})$\;
		\If{$c_l^\star < c_G^\star$}{
			$\Wv_G^\star \leftarrow \Wv_l^\star$ and $c_G^\star \leftarrow c(\Wv_G^\star)$\;
		}
	}
        }
    }
\caption{PSO Algorithm}\label{alg:PSO}
\end{algorithm}
\bigskip 

The PSO algorithm keeps track of local and global best particles throughout the optimization iterations. The local best particles are denoted by $\Wv_l^\star$, $l\in[1:L]$ and the global best value is denoted by  $\Wv_G^\star$. The global best particle will be chosen as the final value at the end of the procedure. The update at each iteration is done based on the local and global best particles by defining a velocity term which is updated by the following procedure,
\begin{align}
\Vv_{l,j} \leftarrow \eta \Vv_{l,j} &+ {a_1}\mathbf{r}_1\circ\left( \Wv^\star_{l,j} - \Wv_{l,j} \right) +{a_2}\mathbf{r}_1\circ\left( \Wv_{G,j}^{\star} - \Wv_{l,j} \right),
\end{align}
where $\circ$ is the Hadamard product, $\Wv^\star_{l,j}$ and $\Wv_{G,j}^{\star} $ are the $j$th row vector of $\Wv^\star_{l}$ and $\Wv_G^\star$, respectively, similar to the relation between $\Wv_{l}$ and $\Wv_{l,j}$ defined in~\eqref{eq:pariticle-def}. For the initialization of the velocity term, $\Vv_{l,j}$ is chosen uniformly at random over $\mathcal{V}$, where
\begin{align*}
\mathcal{V}&=[-|x_{D,\max}-x_{D,\min}|,|x_{D,\max}-x_{D,\min}|]\\
&\quad\times[-|y_{D,\max}-y_{D,\min}|,|y_{D,\max}-y_{D,\min}|]\\
&\quad\times[-|z_{D,\max}-z_{D,\min}|,|z_{D,\max}-z_{D,\min}|].
\end{align*}
The constant $\eta $ is the inertia weight which is used to control the convergence speed, $a_{1}$ and $a_{2}$ are constants representing the step size that the particles takes toward its local best and global best solutions, respectively, and ${\mathbf{r} _1}$ and $\mathbf{r}_2$ are $1 \times 3$ random vectors where each element is chosen i.i.d. from $\U(0,1)$.
Based on the velocity term, the particles are updated by
\begin{equation}
\Wv_{l,j} \leftarrow \Wv_{l,j}+\Vv_{l,j}.
\end{equation}
This process is repeated until the global best solution $\Wv_G^\star$ converges or attains the maximum number of iterations. 

Finally, it remains to explain the cost function $c(\Wv_l)$. In our proposed PSO, 
the cost function $c(\Wv_l)$ itself involves an optimization process and is given by 
\begin{align}
\mathop {\min }\limits_{{b_{k0}},{p_{k0}},{\tau _{kj}}} \sum\limits_{k=1}^K \left({{r_T} - {r_k}\left( {{b_{k0}},{p_{k0}},{\tau _{kj}}|\Wv_l} \right)} \right), \label{new_opti2}
\end{align}
where the minimum is over all parameters subject to
\begin{align}
\sum_{k=1}^K b_{k0} &\le B_{0}, \label{new_opti2_last2} \\ 
\sum_{k=1}^K p_{k0} &\le P_{0},\\
\tau_{kj}b_{k0}\C\left(\sfrac{|h_{j,0}|^2p_{k0}}{b_{k0}N_0}\right) &\ge \tau_{kj}r_{k}\left( {{b_{k0}},{p_{k0}},{\tau _{kj}}|\Wv_l} \right), \nonumber\\
&\forall k, j\\
{r_T} - {{r_k}\left( {{b_{k0}},{p_{k0}},{\tau _{kj}}|\Wv_l} \right)} &\ge 0, \forall k \label{new_opti2_last3}\\
\sum\limits_{k=1}^K {{\tau _{kj}}}  &\le S_D^{(j)}, \forall j \label{new_opti2_last1}
\end{align}
where ${r_k}\left( {{b_{k0}},{p_{k0}},{\tau _{kj}}|\Wv_l} \right)$ is the transmission rate~\eqref{eq:rate} assuming that $\Wv_l$ is fixed.
Since the above problem is a resource allocation problem (including DBS--terminal association) for a fixed set of DBS locations $\Wv_l$, we refer to the above problem as the {\em resource allocation problem (RAP)}.

\begin{lemma}
The RAP is a convex optimization problem on ${b_{k0}}$, ${p_{k0}}$, and ${\tau _{kj}}$.
\end{lemma}
\begin{IEEEproof}
The proof is deferred to Appendix \ref{app2}.
\end{IEEEproof}

Since the RAP is a convex optimization problem, it can be solved by finding the KKT conditions stated in the following proposition. 

\begin{proposition}\label{prop:KKT}
The optimal bandwidth and power allocation for terminal $k$ at the MBS satisfies
\begin{multline}\label{KKT_condition_1}
\left( { - 1 + \sum\limits_{j=1}^N {{\nu _{kj}}} {\tau _{kj}} + {\rho _k}} \right)\left( {C\left( {\frac{{|a_{k,0}|^2{p_{k0}}}}{{{b_{k0}}{N_0}}}} \right) - \frac{{|a_{k,0}|^2{p_{k0}}}}{{{b_{k0}}{N_0} + |a_{k,0}|^2{p_{k0}}}}} \right) + \lambda \\
- \sum\limits_{j=1}^N {{\nu _{kj}}} {\tau _{kj}}\left( {C\left( {\frac{{|h_{j,0}|^2{p_{k0}}}}{{{b_{k0}}{N_0}}}} \right) - \frac{{|h_{j,0}|^2{p_{k0}}}}{{{b_{k0}}{N_0} + |h_{j,0}|^2{p_{k0}}}}} \right) = 0
\end{multline}
\begin{align}\label{KKT_condition_2}
\left( { - 1 + \sum\limits_{j=1}^N {{\nu _{kj}}} {\tau _{kj}} + {\rho _k}} \right)\left( {\frac{{|a_{k,0}|^2{b_{k0}}}}{{{b_{k0}}{N_0} + |a_{k,0}|^2{p_{k0}}}}} \right) + \mu  - \sum\limits_{j=1}^N {{\nu _{kj}}} {\tau _{kj}}\left( {\frac{{|h_{j,0}|^2{b_{k0}}}}{{{b_{k0}}{N_0} + |h_{j,0}|^2{p_{k0}}}}} \right) = 0
\end{align}
and the optimal time faction allocation for terminal $k$ at DBS $j$ satisfies
\begin{align} \label{eq:tau_sol}
{\tau _{kj}} = \frac{1}{{2bC\left( {\frac{{|g_{k,j}|^2p}}{{b{N_0}}}} \right)}}{\left[ {\frac{{bC\left( {\frac{{|g_{k,j}|^2p}}{{b{N_0}}}} \right)\left( {1 - {\rho _k}} \right) - {\theta _j}}}{{{\nu _{kj}}}} + {b_{k0}}\left( {C\left( {\frac{{|h_{j,0}|^2{p_{k0}}}}{{{b_{k0}}{N_0}}}} \right) - C\left( {\frac{{|a_{k,0}|^2{p_{k0}}}}{{{b_{k0}}{N_0}}}} \right)} \right)} \right]^ + }
\end{align}
where ${\left[ a \right]^ + } = \max \left( {a,0} \right)$ and $\lambda$, $\mu$, $\nu _{kj}$, $\rho _k$, and ${{\theta _j}}$ are the Lagrange multipliers corresponding to the constaints in RAP.
\end{proposition}
\begin{IEEEproof}
The proof is deferred to Appendix \ref{app:KKT}.
\end{IEEEproof}

Finally, the optimal resource allocation parameters can be found by solving the system of equations~\eqref{KKT_condition_1},~\eqref{KKT_condition_2} along with~\eqref{eq:tau_sol} where the explicit steps are given in Appendix~\ref{app:KKT}.

\section{Numerical Evaluations} \label{sec:numerical_evaluation}
\begin{table}[!t]
\renewcommand{\arraystretch}{1.3}
\caption{Simulation parameters}
\label{table_1}
\centering
\begin{tabular}{c|c|c}
\hline
\bfseries Symbol & \bfseries Definition & \bfseries Value\\
\hline\hline
${B_{0}}$ & Total bandwidth of the MBS & ~20 MHz \\
\hline
${b}$ & Bandwidth of each DBS &~5 MHz \\
\hline
${P_{0}}$ &Total transmit power of the MBS &  ~20 W\\
\hline
${p}$ & Transmit power of each DBS & ~5 W\\
\hline
${N}$ & Number of terminals & ~10\\
\hline
${\eta }$ & Inertia weight & ~0.7298\\
\hline
${a_{1}, a_{2}}$ & Personal and social acceleration coefficients & ~1.4962 \\
\hline
${\Omega }$ &CW size of Wi-Fi APs& ~16 \\
\hline
${m}$ & Maximum backoff stage of Wi-Fi APs& ~3 \\
\hline
\end{tabular}
\end{table}

\begin{table}[!t]
\renewcommand{\arraystretch}{2.2}
\caption{Urban environment parameters \cite{ Air-to-Ground-channel}}
\label{table_2}
\centering
\begin{tabular}{c|c|c}
\hline
\bfseries Symbol & \bfseries Definition & \bfseries Value\\
\hline\hline
$\alpha, \beta$ & \parbox{14em}{Coefficients that depend on the environment}   & ~9.61, 0.16 \\
\hline
${L_{\operatorname{LoS}}}, {L_{\operatorname{NLoS}}}$ & \parbox{14em}{Air-to-ground pathloss terms for LoS and NLoS components} & ~1, 20 dB \\
\hline
\end{tabular}
\end{table}

In this section, we perform numerical evaluations of the proposed optimization strategy. 
In the simulations, we consider a single MBS cell in which multiple DBSs and Wi-Fi APs coexist.
We assume that the MBS is located at the center of the 2D network area and each terminal is uniformly located at random over the 2D network area of $[-1,1]\times[-1,1]$ $\text{km}^2$, i.e., $\phi_0=(0,0,0)$ and $\mathcal{T}=[-1,1]\times [-1,1]\times \{0\}$ $\text{km}^3$.
We assume that  each DBS is located in the same 2D network area but its altitude is in between $0.1$ and $0.4$ $\text{km}$, i.e., $\mathcal{D}=[-1,1]\times [-1,1]\times [0.1,0.4]$ $\text{km}^3$.
The simulation parameters and the parameters involving the urban environment are summarized in Tables \ref{table_1} and \ref{table_2}.

\begin{figure}[!t]
\centering
\includegraphics[width=3.2in]{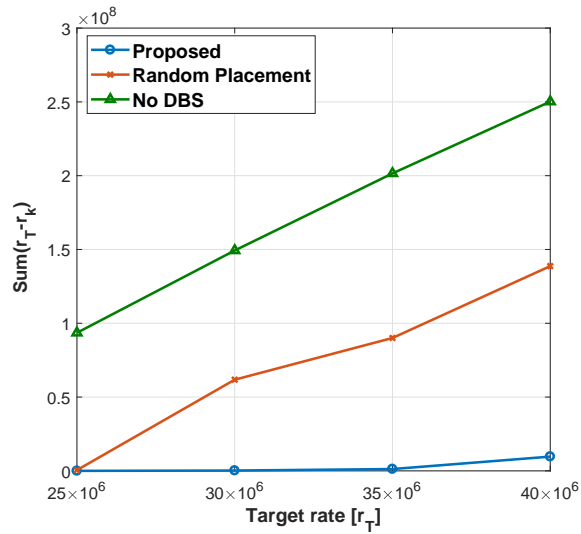}
\caption{Performance comparison between the proposed scheme and the baseline schemes with respect to the target rate of the terminals.}
\label{simul_fig1}
\end{figure}
Figure \ref{simul_fig1} shows the aggregate gaps between the communication rate and the target rate of the terminals for a target rate $r_T$ when $N=3$, $S_D^{(j)}=0.6$. The parameter $S_D^{(j)}$, i.e., the active time fraction for DBS $j$, is chosen as in Remark~\ref{rmk:cw} with the Wi-Fi AP's CW size $\Omega=16$ and the maximum backoff stage $m=3$. 
The performance of the proposed scheme is compared with the performance of two baseline strategies which we will refer to as the `Random DBS placement scheme' and the `No DBS scheme'. 
As the names suggest, the Random DBS placement scheme optimizes the radio resource allocation of MBS and DBSs by assuming {\em random positions} of the DBSs uniformly distributed over $\mathcal{D}$ (without updating the positions of the DBSs). The `No DBS scheme' only optimizes the radio resource allocation of MBS, i.e., it does not utilize any DBSs.
As illustrated in Fig. \ref{simul_fig1}, the proposed scheme performs better than the Random DBS placement and the No DBS schemes. The figure demonstrates the benefits of utilizing DBSs as well as the benefits of optimizing the 3D positions of the DBSs. Also, as the target rate of the terminals increases, the performance difference between the proposed and the compared schemes also increases. This means that the proposed scheme could be utilized more efficiently in the cases where most terminals in the coverage of the MBS require wireless services with high data rates.  

\begin{figure}[!t]
\centering
\includegraphics[width=3.2in]{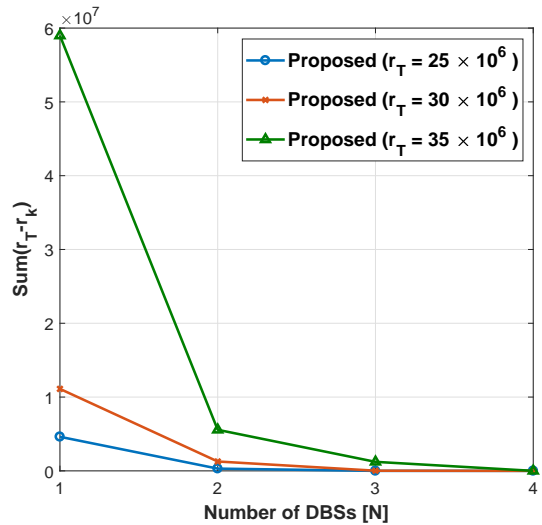}
\caption{Plot of the aggregate gap achieved through the proposed scheme vs. the number of DBSs.}
\label{simul_fig2}
\end{figure}

Next, Fig. \ref{simul_fig2} illustrates the aggregate gap between the communication rates and the target rate of terminals with respect to the change in the number of utilized DBSs. 
One important observation we can make from Fig. \ref{simul_fig2} is that it identifies the minimum number of DBSs that is need for all the terminals to satisfy the target rate. 
This implies that through a parallel execution of the proposed algorithm with different number of DBSs, we can find the minimum number of DBSs for which all terminals satisfy the required target rate.

\begin{figure}[!t]
\centering
\includegraphics[width=3.2in]{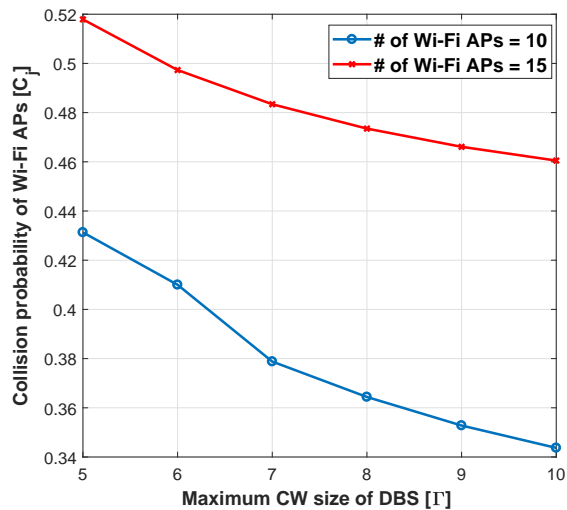}
\caption{Plot of the collision probability of Wi-Fi APs vs. the maximum CW size of the DBSs.}
\label{simul_fig3}
\end{figure}

Figure \ref{simul_fig3} presents the collision probabilities of the Wi-Fi APs with respect to the maximum CW size of the DBSs, which can be obtained through solving the nonlinear systems of equations \eqref{access_wifi} to \eqref{collision_DBS} as discussed in Remark~\ref{rmk:cw}. 
From Fig. \ref{simul_fig3}, we can find the optimal CW size of the DBSs given the collision probability of the Wi-Fi APs. The optimal CW size of the DBSs is set to the minimum value satisfying the required collision probability of the Wi-Fi APs. This is because, as the value of the CW size of the DBSs decreases, the performance of the DBSs improves whereas the performance of the Wi-Fi APs deteriorates. For example, when the required collision probability of $10$ Wi-Fi APs is $0.5$, the optimal CW size of the DBSs is given by $6$. Note that the derived CW size of the DBSs under the considered network setting is used to determine $S_D^{(j)}$ via (\ref{time_share}).

\begin{figure}[!t]
\centering
\includegraphics[width=3.2in]{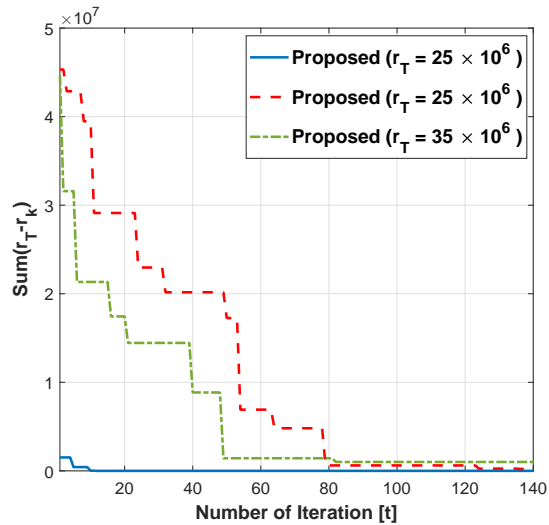}
\caption{Plot of the aggregate gap achieved through the proposed scheme vs. number of iterations $t$.}
\label{simul_fig4}
\end{figure}

\begin{figure*}[th]
\centering
\subfigure[$\phi_{j,x}$-axis]{
\includegraphics[width=0.31\linewidth]{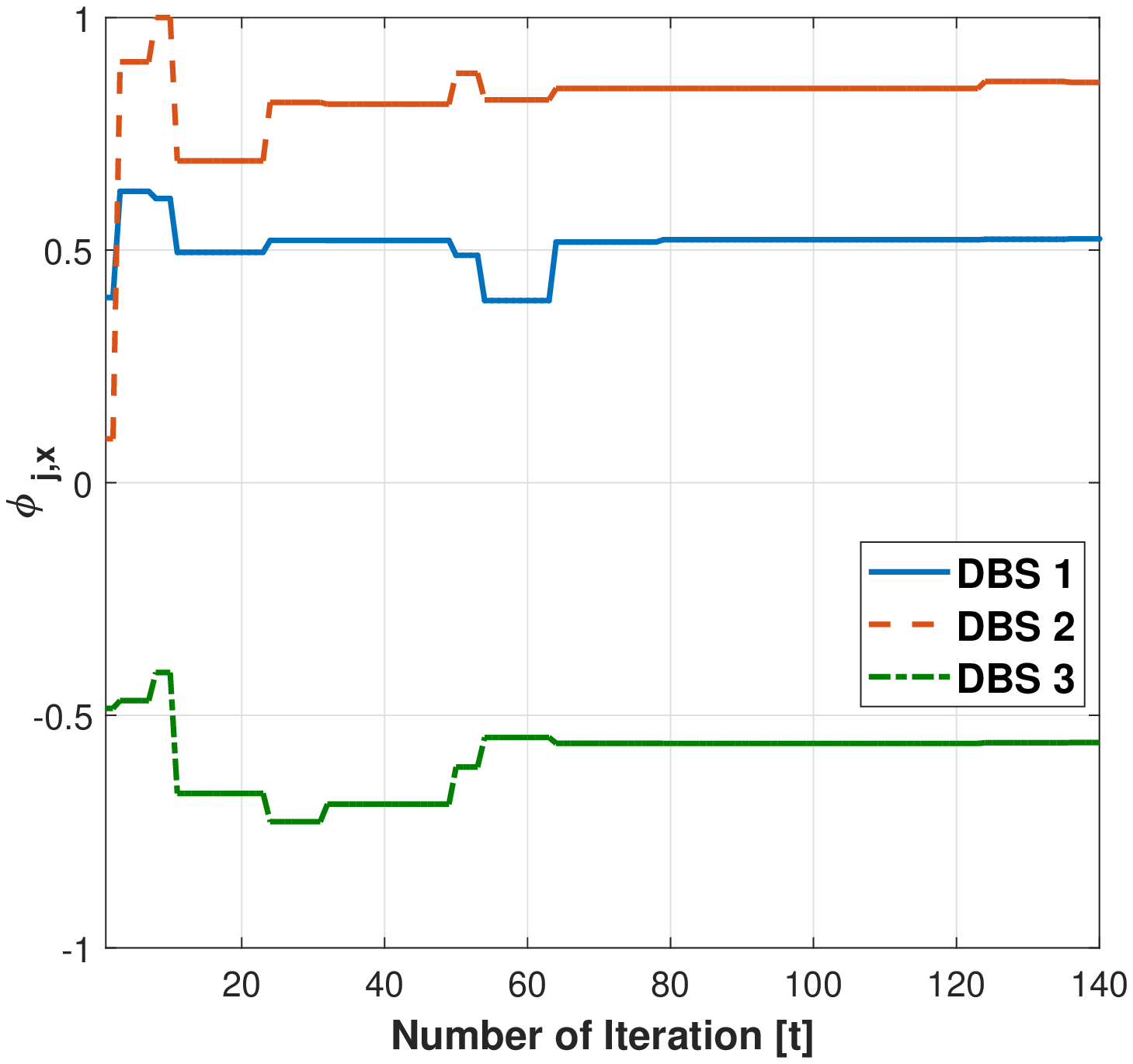}
}
\centering
\subfigure[$\phi_{j,y}$-axis]{
\includegraphics[width=0.31\linewidth]{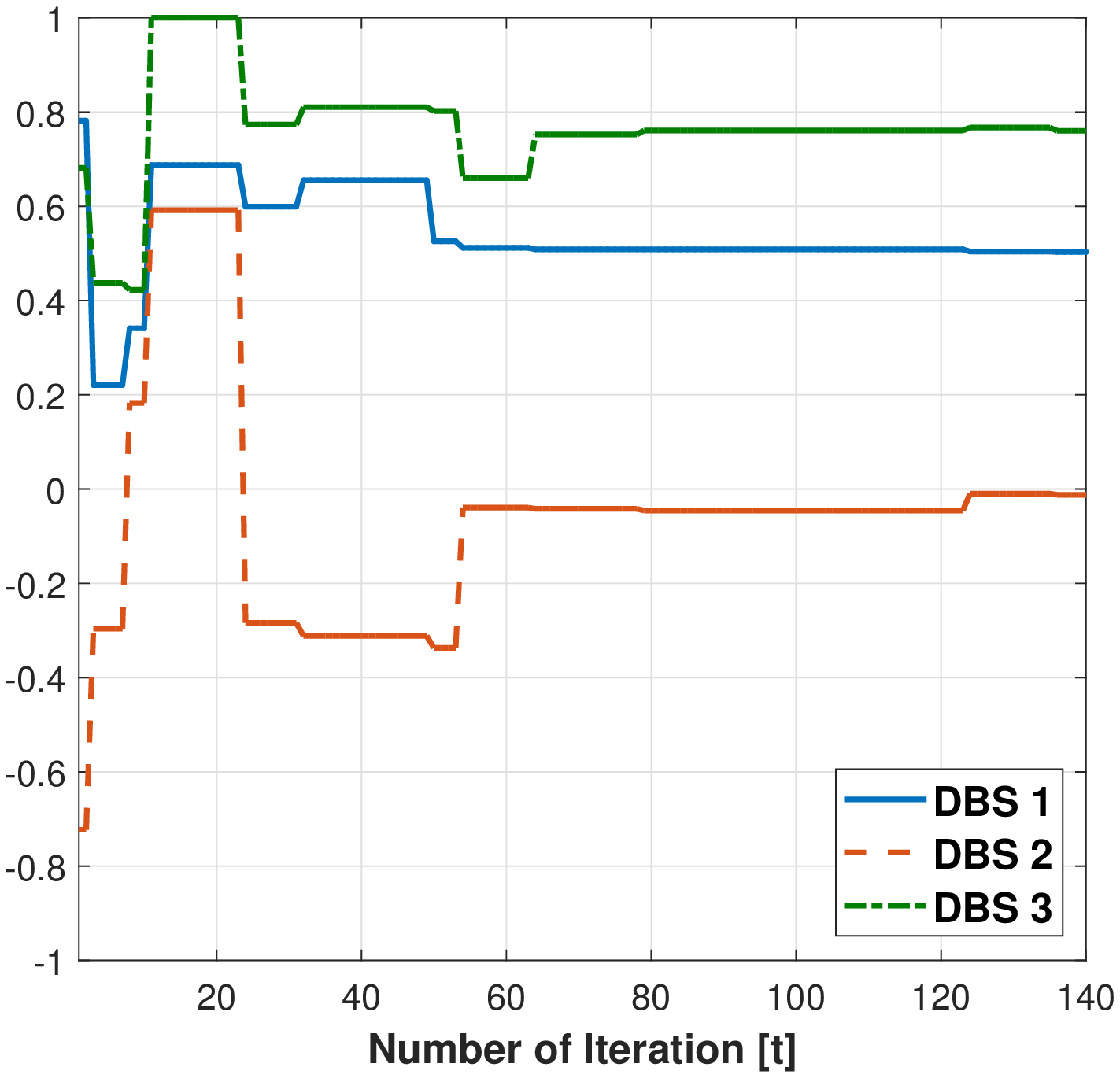}
}
\centering
\subfigure[$\phi_{j,z}$-axis]{
\includegraphics[width=0.31\linewidth]{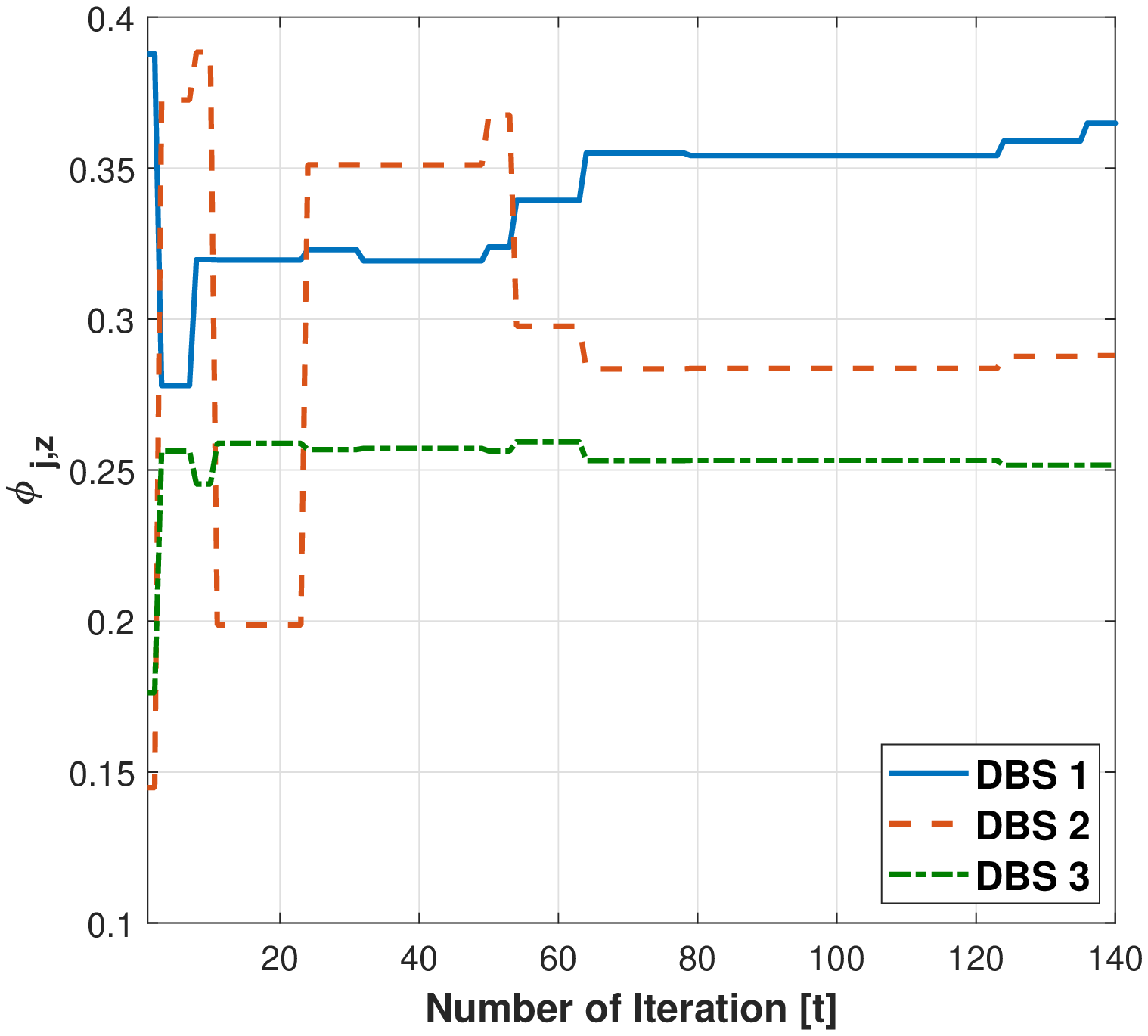}
}
\caption{A plot of the 3D position trajectories of the DBSs vs. iteration $t$.} \label{simul_fig5}
\end{figure*}

\begin{figure}[!t]
\centering
\includegraphics[width=3.2in]{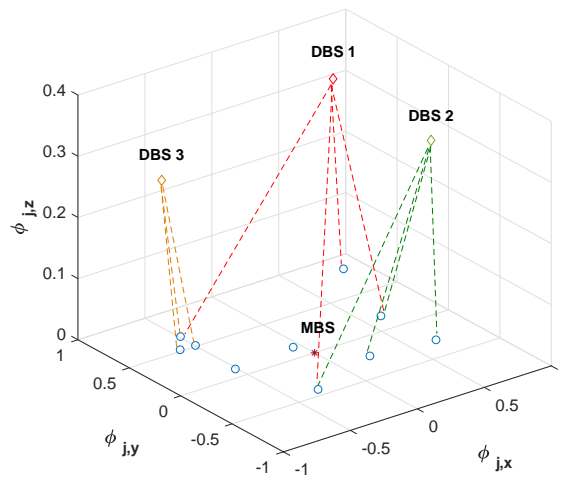}
\caption{Plot of an example instance of the optimized DBS 3D positions and the DBS--terminal associations. Note that the proposed scheme suggests different optimal elevations of the DBSs depending on the terminal distributions and that each DBS assists multiple terminals.}
\label{simul_fig6}
\end{figure}

In Fig. \ref{simul_fig4}, we show the convergence process of the proposed scheme by observing the aggregate rate gap with respect to the number of iterations $t$. As the iteration progresses, the aggregate gap becomes smaller and finally converges to a specific value. In Fig. \ref{simul_fig5}, we show the convergence trajectory of the 3D coordinates with respect to the number of iterations. Lastly, Fig. \ref{simul_fig6} illustrates an example instance of the optimized 3D positions of the DBSs and the associations (which DBS assists which terminal, depicted as dotted connections).

\section{Concluding Remarks}\label{sec: concluding_remark}
Looking back, we recognize several distinctions between a mobile DBS assisted network and conventional (fixed and grounded) small base station assisted networks. Most distinctively, DBS assisted networks do not have high throughput backhaul links as in most conventional small base station assisted networks. This entails several complications. Firstly, the backhaul link can no longer be modeled as a wireline link with some fixed reliable throughput, and instead, DBSs are connected to their source nodes via wireless links. Such wireless links are in many cases different from high capacity backhaul links with line-of-sight beamforming since the DBSs are themselves mobile and in some sense temporary. Such distinction naturally suggests that the DBSs should be viewed as relays, i.e., one would need to develop a strategy by considering both hops jointly. 

Another distinction from conventional small cell networks is that the DBSs are mobile. Thus, the cell planning aspect of optimizing a network becomes much more complicated and dynamic. 
Moreover, the elevation of the DBS location becomes another important parameter for optimizing the network. 

Our work is an (early) attempt in understanding the fundamental behavior of DBS aided networks while considering both distinctions. In particular, we formulate a joint optimization problem to find the best resource allocations for both hops simultaneously (MBS and DBS resource allocations), and to find the optimal DBS locations.

Following our point of view, it would be interesting future work to consider and compare the performance of other sophisticated relaying strategies beyond DF, e.g., compress--forward~\cite{Cover--El-Gamal1979, Lim--Kim--El-Gamal--Chung2011}, compute--forward~\cite{Nazer--Gastpar2011}, as well as amplify--forward~\cite{Schein--Gallager2000} based strategies. Another interesting direction would be to consider a dynamic model where the location optimization varies with time while considering the current locations.

\appendices
\section{Proof of Theorem~\ref{thm:DF}} \label{app:proof-thm1}
The DF strategy presented in~\cite{Cover--El-Gamal1979, El-Gamal--Kim2011} can be easily extended to the multiple relay case presented by the following proposition.

\begin{proposition}
A rate tuple $(r_1,\ldots, r_{K})$ is achievable by the cooperative multi-DBS DF strategy if
\begin{align}
r_k &\leq  I(X_{k,0};Y_{k,0})+\sum_{j=1}^{N} \tau_{kj}I(X^{\operatorname{DBS}}_{k,j};Y_{k,j}),\label{eq:constapp1}\\
r_k &\leq I(X_{k,0}; Y^{\operatorname{DBS}}_{j,k}), \quad j\in[1:N] \text{ such that } \tau_{kj}> 0. \label{eq:constapp2}
\end{align}
\end{proposition}
Note that each mutual information can be seen as a point-to-point capacity of a band-limited Gaussian channel. Since for an AWGN channel with bandwidth $W$, power constraint $P$, and AWGN with $\Cc\Nc(0, N_0)$, the capacity~\cite{Wyner1966, Gallager1968} is given as
\begin{align} \label{eq:shannon_cap}
W\log\left(1+\frac{P}{N_0 W}\right),
\end{align}
by  substituting \eqref{eq:shannon_cap} into \eqref{eq:constapp1} and \eqref{eq:constapp2}, we have
\begin{align}
r_k &\leq b_{k0}\C\left(\sfrac{|a_{k,0}|^2p_{k0}}{b_{k0}N_0}\right) +  \sum_{j=1}^{N}\tau_{kj}b\C\left(\sfrac{|g_{k,j}|^2p}{bN_0}\right),
\end{align}
\begin{align}
&r_k \leq b_{k0}\C\left(\sfrac{|h_{j,k}|^2p_{k0}}{b_{k0}N_0}\right), \quad j\in[1:N] \nonumber\\
&\text{ such that } \tau_{kj}> 0. \label{eq:constapp3}
\end{align}
Finally, by the continuity of $\C(\cdot)$ and by rewriting the constraint~\eqref{eq:constapp3} as
\begin{align}
\tau_{kj} r_k &\le \tau_{kj} b_{k0}\C\left(\sfrac{|h_{j,k}|^2p_{k0}}{b_{k0}N_0}\right), \quad j\in[1:N]
\end{align}
the rate constraints \eqref{eq:DF-rate} and \eqref{eq:DF-rate2} are obtained, which completes the proof of Theorem~\ref{thm:DF}.

\section{Convexity of the RAP}\label{app2}
To Identify the convexity of the objective function (\ref{eq:objective}), we first check some properties of ${r_k}$ in the objective function. Let us present ${r_k}$ as the following general form:
\begin{align}
r(c_1,c_2,c_3 ) = c_1\ln \left( {1 + \frac{c_2}{c_1}} \right) + c_3.
\end{align}
Then, the convexity of the function $r(b,p,\tau)$ can be checked by first looking at its first partial derivatives given by
\begin{equation}
\nabla r = \left( {\begin{array}{*{20}{c}}
{\frac{{\partial f}}{{\partial b}}}\\
{\frac{{\partial f}}{{\partial p}}}\\
{\frac{{\partial f}}{{\partial \tau }}}
\end{array}} \right) = \left( {\begin{array}{*{20}{c}}
{\ln \left( {1 + \frac{p}{b}} \right) - \frac{p}{{b + p}}}\\
{\frac{b}{{b + p}}}\\
1
\end{array}} \right).
\end{equation}
Then, the Hessian of the function $r(b,p,\tau)$ is evaluated by its second order partial derivatives,
\begin{align}
H &= \left( {\begin{array}{*{20}{c}}
{\frac{{{\partial ^2}f}}{{{\partial ^2}{b^2}}}}&{\frac{{{\partial ^2}f}}{{\partial b\partial p}}}&{\frac{{{\partial ^2}f}}{{\partial b\partial \tau }}}\\
{\frac{{{\partial ^2}f}}{{\partial p\partial b}}}&{\frac{{{\partial ^2}f}}{{{\partial ^2}{p^2}}}}&{\frac{{{\partial ^2}f}}{{\partial p\partial \tau }}}\\
{\frac{{{\partial ^2}f}}{{\partial \tau \partial b}}}&{\frac{{{\partial ^2}f}}{{\partial \tau \partial p}}}&{\frac{{{\partial ^2}f}}{{{\partial ^2}{\tau ^2}}}}
\end{array}} \right) \\ 
&= \left( {\begin{array}{*{20}{c}}
{\frac{p}{{b + p}}\left( {\frac{p}{{b + p}} - \frac{1}{b}} \right)}&{\frac{p}{{{{\left( {b + p} \right)}^2}}}}&0\\
{\frac{1}{{b + p}}\left( {1 - \frac{b}{{b + p}}} \right)}&{\frac{{ - b}}{{{{\left( {b + p} \right)}^2}}}}&0\\
0&0&0
\end{array}} \right).
\end{align}
Based on the Hessian, the leading principal minors of $r(b,p,\tau)$ can be computed as follows
\begin{align}
D_1 &= \frac{p}{{b + p}}\left( {\frac{p}{{b + p}} - \frac{1}{b}} \right) \le 0, \label{minor1}\\
D_2 &= \det \left( {\begin{array}{*{20}{c}}
{\frac{{{\partial ^2}f}}{{{\partial ^2}{b^2}}}}&{\frac{{{\partial ^2}f}}{{\partial b\partial p}}}\\
{\frac{{{\partial ^2}f}}{{\partial p\partial b}}}&{\frac{{{\partial ^2}f}}{{{\partial ^2}{p^2}}}}
\end{array}} \right) \nonumber\\
&= \frac{p}{{{{\left( {b + p} \right)}^3}}}\left( {1 - \frac{b}{{b + p}} - 1 + \frac{b}{{b + p}}} \right) = 0, \label{minor2}\\
{D_3} &= \det \left( H \right) = 0. \label{minor3}
\end{align}
The results of (\ref{minor1}) - (\ref{minor3}) are obtained under the condition of $b$, $p$, and $\tau$ $\ge 0$. 
It is identified that the signs of the first, second, and third leading pricipal monors are non-positive, 0, and 0, repectively. Thus, the Hessian of $r(b,p,\tau)$ is negative semidefinite. This means that the function $r(b,p,\tau)$ is a concave function, such that $-r(b,p,\tau)$ is a convex function. 
The objective function (\ref{eq:objective}) is a positive linear combination of $-r(b,p,\tau)$ type functions. Thus, we conclude that the objective function (\ref{eq:objective}) is a convex function.  

\section{}\label{app:KKT}
The Lagrangian associated to the optimization problem RAP can be presented as
\begin{align}
&L\left( {{b_{k0}},{p_{k0}},{\tau _{kj}},\lambda ,\mu ,{\nu _{kj}},{\rho _k},{\theta _j}} \right) \nonumber \\
&= \sum\limits_k {{r_T} - {r_k}\left( {{b_{k0}},{p_{k0}},{\tau _{kj}}} \right)} \nonumber\\ 
&+ \lambda \left( {\sum\limits_k {{b_{k0}}}  - {B_0}} \right) + \mu \left( {\sum\limits_k {{p_{k0}}}  - {P_0}} \right)\nonumber \\
&+{\sum\limits_j {{\nu _{kj}}{\tau _{kj}}\left( {{r_k}\left( {{b_{k0}},{p_{k0}},{\tau _{kj}}} \right) - {b_{k0}}C\left( {\frac{{|h_{j,0}|^2{p_{k0}}}}{{{b_{k0}}{N_0}}}} \right)} \right)} } \nonumber\\
&+ \sum\limits_k {{\rho _k}\left( {{r_k}\left( {{b_{k0}},{p_{k0}},{\tau _{kj}}} \right) - {r_T}} \right) } \nonumber\\
&+ \sum\limits_j {{\theta _j}\left( {\sum\limits_{k } {{\tau _{kj}}}  - S_D^{(j)}} \right)} ,  
\end{align}
where $\lambda$, $\mu$, ${\nu _{kj}}$, ${\rho _k}$, and ${\theta _j}$ are the Lagrange multipliers corresponding to the constraints \eqref{new_opti2_last2} to \eqref{new_opti2_last1}.
Also, the Lagrange dual function is presented as 
\begin{multline}\label{dual_func2}
D\left( {\lambda ,\mu ,{\nu _{kj}},{\rho _k},{\theta _j}} \right) \\
 = \mathop {\min }\limits_{{b_{k0}},{p_{k0}},{\tau _{kj}}} L\left( {{b_{k0}},{p_{k0}},{\tau _{kj}},\lambda ,\mu ,{\nu _{kj}},{\rho _k},{\theta _j}} \right),
\end{multline}
and the dual problem corresponding to the primal problem is given as 
\begin{align}\label{dual_pro1}
\mathop {\max }\limits_{\lambda ,\mu ,{\nu _{kj}},{\rho _k},{\theta _j}} D\left( {\lambda ,\mu ,{\nu _k},{\rho _k},{\theta _{kj}}} \right).
\end{align}
By the KKT condition, an optimal solution of the optimization problem \eqref{new_opti2} to \eqref{new_opti2_last1} must satisfy the following conditions:
\begin{align}
\frac{{\partial L}}{{\partial {b_{k0}}}} &= 0, \forall k \\
\frac{{\partial L}}{{\partial {b_{k0}}}} &= 0, \forall k \\
\frac{{\partial L}}{{\partial {\tau _{kj}}}} &= 0,\forall k,j.
\end{align}
Thus, for the given Lagrange multipliers, the optimal bandwidth and power allocation for terminal $k$ at the MBS satisfies ~\eqref{KKT_condition_1}~-~\eqref{KKT_condition_2}, and the optimal time fraction allocation for terminal $k$ at DBS $j$ satisfies~\eqref{eq:tau_sol}.

The update of the Lagrange multipliers for achieving the optimal solution is performed by solving the dual problem (\ref{dual_pro1}). For given ${{b _{k0}}}$, ${{p _{k0}}}$, and ${{\tau _{kj}}}$, the dual problem \eqref{dual_pro1} can be simplified as 
\begin{align}
&\mathop {\max }\limits_{\lambda  \ge 0} \lambda \left( {\sum\limits_k {{b_{k0}}}  - {B_0}} \right) + \mathop {\max }\limits_{\mu  \ge 0} \mu \left( {\sum\limits_k {{p_{k0}}}  - {P_0}} \right) \nonumber \\
&+ \sum\limits_k {\sum\limits_j {\mathop {\max }\limits_{{\nu _{kj}} \ge 0} {\nu _{kj}}{\tau _{kj}}} } ({r_k}\left( {{b_{k0}},{p_{k0}},{\tau _{kj}}} \right)\nonumber\\
&\,\,\,\,\,\,\,\,\,\,\,\,\,\,\,\,\,\,\,\,\,\,\,\,\,\,\,\,\,\,\,\,\,\,\,\,\,\,\,\,- {b_{k0}}C(\frac{{|{h_{j,0}}{|^2}{p_{k0}}}}{{{b_{k0}}{N_0}}})) \nonumber\\
&+ \sum\limits_k {\mathop {\max }\limits_{{\rho _k} \ge 0} {\rho _k}\left( {{r_k}\left( {{b_{k0}},{p_{k0}},{\tau _{kj}}} \right) - {r_T}} \right) } \nonumber\\
&+\sum\limits_j {\mathop {\max }\limits_{{\theta _j} \ge 0} {\theta _j}\left( {\sum\limits_{k} {{\tau _{kj}}}  - S_D^{(j)}} \right)}.
\end{align}
For a differentiable dual function, the gradient-descent method can be applied to update the Lagrange multipliers by
\begin{align}
{\lambda ^{t + 1}} &= {\left[ {{\lambda ^t} + {\zeta _1}\left( {\sum\limits_{k} {{b_{k0}}}  - {B_0}} \right)} \right]^ + }, \nonumber \\
{\mu ^{t + 1}} &= {\left[ {{\mu ^t} + {\zeta _2}\left( {\sum\limits_{k} {{p_{k0}}}  - {P_0}} \right)} \right]^ + }, \nonumber \\
\nu _{kj}^{t + 1} &= [\nu _k^t + {\zeta _3}{\tau _{kj}}({r_k}({b_{k0}},{p_{k0}},{\tau _{kj}}) \nonumber \\
& \,\,\,\,\,\,\,\,\,\,\,\,\,\,\,\,- {b_{k0}}C(\frac{{|{h_{j,0}}{|^2}{p_{k0}}}}{{{b_{k0}}{N_0}}})){]^ + } \nonumber\\
\rho _k^{t + 1} &= {\left[ {\rho _k^t + {\zeta _4}\left( {{r_k}\left( {{b_{k0}},{p_{k0}},{\tau _{kj}}} \right) - {r_T}} \right)} \right]^ + }, \nonumber\\
\theta _j^{t + 1}{\rm{ }} &= {\left[ {\theta _j^t + {\zeta _5}\left( {\sum\limits_{k \in U} {{\tau _{kj}}}  - S_D^{(j)}} \right)} \right]^ + },
\end{align}    
where ${{\zeta _1}}$ to ${{\zeta _5}}$ are the step sizes for which the small constant value should be chosen to converge the Lagrange multipliers.


\newcommand{\noopsort}[1]{}

\end{document}